# Survival May Not be for the Fittest (Lessons from some TV games)

E.Ahmed and A.S.Hegazi
Mathematics Department, Faculty of Sciences, Mansoura 35516, EGYPT.

**Abstract:**
In this paper we argue that biological fitness is a multi-objective concept hence the statement "fittest" is inappropriate. The following statement is proposed "Survival is mostly for those with non-dominated fitness". Also we use some TV games to show that under the following conditions:
  i)   There are no dominant players.
  ii)  At each time step successful players may eliminate some of their less successful competitors,
Then the ultimate winner may not be the fittest (but close).

1. Multi-objective fitness:
Definition (1): A measure of biological fitness of an animal is the number of its offspring that reaches maturity.
An equivalent definition is that it is the number of its offspring that are in the best possible state (to ensure that they will have their offspring).
It is clear that this definition is a multi-objective one since it can be decomposed into the following two objectives:
  i)   Maximize the number of offspring.
  ii)  Maximize their state (e.g. food, protection, training to hunt etc…).
These objectives are contradictory hence biological fitness is a multi-objective concept. Now we recall some basics of multi-objective optimization [Ehrgott 2005]. Almost every real life problem is multi-objective (MOB). Methods for MOB optimization are mostly intuitive.
**Definition (2):** A MOB problem is:
Minimize (min) $Z_i(\underline{x}), i=1,2,...,k$ subject to $\underline{g}(\underline{x}) \leq \underline{0}, \underline{h}(\underline{x}) = \underline{0}$   (1)
**Definition (3):** A vector $\underline{x}^*$ dominates $\underline{x}'$ if $Z_i(\underline{x}) \leq Z_i(\underline{x}') \forall i=1,2,...,k$ with strict inequality for at least one i, given that all constraints are satisfied for both vectors.

A non-dominated solution $\underline{x}^*$ is called Pareto optimal and the corresponding vector $Z_i(\underline{x}^*), i=1,2,...,k$ is called efficient. The set of such solutions is called a Pareto set.
Now we discuss some methods for solving MOB problems:
The first method is the lexicographic method. In this method objectives are ordered according to their importance. Then a single objective problem is solved while completing the problem gradually with constraints i.e.
$\min Z_1$ subject to
$\underline{g}(\underline{x}) \leq \underline{0}, \ \underline{h}(\underline{x}) = \underline{0}$   (2)
then if ZMIN(1) is the solution the second step is
$\min Z_2$ subject to $Z_1 = ZMIN(1)$, and the constraints in (2)
and so on.
A famous application is in university admittance where students with highest grades are allowed in any college they choose. The second best group is allowed only the remaining places and so on. This method is useful but in some cases it is not applicable.

**Proposition (1):** An optimal solution for the lexicographic problem is Pareto optimal.
Proof: Let $\underline{x}^*$ be the solution to the Lexicographic problem $P_l$. Thus $\underline{x} \neq \underline{x}^*$ then $Z_i(\underline{x}) = Z_i(\underline{x}^*), i = 1,2,...,l-1$ and $Z_l(\underline{x}^*) < Z_l(\underline{x})$. Thus $\underline{x}^*$ is not dominated. Q.E.D.

The second method is the method of weights. Assume that it is required to minimize the objectives Z(j), j=1,2,..,n. (The problem of maximization is obtained via replacing Z(j) by -Z(j)). Define

$$Z = \sum_{i=1}^{k} Z_i w(i), \quad 0 \leq w(i) \leq 1, \quad \sum_{i=1}^{k} w(i) = 1 \qquad (3)$$

Then the problem becomes to minimize Z subject to the constraints. This method is easy to implement but it has several weaknesses. The first is that it is not applicable if the feasible set is not convex. The second difficulty of this method is that it is difficult to apply for large number of objectives. However it is quite effective for multi-objective problems with discrete parameters since in this case Pareto optimal set is discrete not a continuous curve.

The third method is the compromise method (sometimes called $\varepsilon$- constraint method $P_\varepsilon(k)$. In this case one minimizes only one objective while setting the other objectives as constraints e.g. minimize Z(k) subject to Z(j)≤a(j) j=2,3,…k-1,k+1,…,n, where a(j) are parameters to be gradually decreased till no solution is found. The problem with this method is the choice of the thresholds a(j). If the solution is unique, then this method is guaranteed to give a Pareto optimal solution.

**Proposition (2):** If the solution is unique, then the $\varepsilon$- constraint method is guaranteed to give a Pareto optimal solution.
Proof: Let $\underline{x}^*$ be the optimal solution for the $\varepsilon$- constraint method then $\forall \underline{x} \neq \underline{x}^* then Z_k(\underline{x}^*) < Z_k(\underline{x})$ hence $\underline{x}^*$ is Pareto optimal. If $\underline{x}^*$ is not unique then it is weakly Pareto i.e. there is no $\underline{x} \neq \underline{x}^*$ such that $Z_i(\underline{x}^*) < Z_i(\underline{x}) \forall i = 1,2,...,n$  Q.E.D

A fourth method using fuzzy logic is to study each objective individually and find its maximum and minimum say ZMAX(j), ZMIN(j) respectively. Then determine a membership m(j)=(ZMAX(j)-Z(j))/(ZMAX(j)-ZMIN(j)). Thus $0 \leq m(j) \leq 1$. Then apply Max{MIN{m(j), j=1,2,…,n}}. Again this method is guaranteed to give a Pareto optimal solution provided that the solution is unique otherwise it is weakly Pareto.

It is clear that in multi-objective problems there is no "best" solution, only non-dominated ones. Hence the term fittest is in-appropriate. We propose to replace it by "non-dominated fitness". This may explain the survival of unfit animals e.g. the Panda (which eats only one type of food and mates only one day per year). Also this answers the question: Which is the fittest? The strong (e.g. lion), the fast (e.g. the cheetah) or the adaptable (e.g. the rat)?

## 2. Games where successful players may eliminate some of their less successful competitors:

In some TV games, at each time step t=1,2,3,…, the players compete. Then the more successful players choose one of the worst two (or three) players and eliminate him (her). It has been observed that in several cases when one of the overall strong players fall into the worst two (or three) players the other competitors chose to eliminate him (her). The logic is that they have a better chance of winning the game if their final opponent is an overall weak. This contradicts the statement "survival for the fittest". In the following we will analyze this game using prisoner's dilemma (PD) game [Hofbauer and Sigmund 1998, Webb 2007].

Game theory is the study of the ways in which strategic interactions among rational players produce outcomes (profits) with respect to the preferences of the players. It consists of three sets: The set of players, set of strategies and the set of payoffs. Each player in a game faces a choice among two or more possible strategies. A strategy is a predetermined program of play that tells the player what actions to take in response to every possible strategy other players may use.
A basic property of game theory is that one's payoff depends on the others' decisions as well as his.

The mathematical framework of the non-cooperative game theory was initiated by von Neumann and Morgenstern in 1944 . Also they had suggested the max-min solution for games which is calculated as follows: Consider two players A and B are playing against each other. Two strategies (S1,S2) (S'1, S'2) are allowed for both of them. This game is called two-player, two-strategy game. Let the payoff matrix be $\begin{bmatrix} a & b \\ d & c \end{bmatrix}$. For zero sum games the max-min solution of von Neumann and Morgenstern is for the first player to choose max{min(a,b), min(c,d)}. The second player chooses min{max(a,c), max(b,d)}. If both quantities are equal then the game is stable. Otherwise use mixed strategies.

A weakness of this formalism has been pointed out by Maynard Smith in the hawk-dove (HD) game whose payoff matrix is 
$\begin{array}{c|cc} & H & D \\ \hline H & (v-c)/2 & v \\ D & 0 & v/2 \end{array}$
.The max-min implies (for v<c) that the solution is D yet as he pointed out this solution is unstable since if one of the players adopts H in a population of D he will a very large payoff which will make other players switch to H and so on till number of H is large enough that they play each other frequently and get the low payoff (v-c)/2. Thus the stable solution is that the fraction of hawks should be nonzero. To quantify this concept one may use the replicator equation which intuitively means that the rate of change of the fraction of players adopting strategy i is proportional to the difference between their payoff and the average payoff of the population i.e.

$$dx_i / dt = x_i[(\Pi x)_i - x\Pi x], i = 1,2,..,n, \sum_{i=1}^{n} x_i = 1 \quad (4)$$

where $x_i$ is the fraction of players adopting strategy i, and $\Pi$ is the payoff matrix. Applying (4) to the HD game one gets that the asymptotically stable equilibrium solution is x=v/c where x is the fraction of hawks in the population.
For asymmetric game the replicator dynamics equation is
$$dx_i / dt = x_i[(\Pi 1 y)_i - x\Pi 1 y], \quad dy_i / dt = y_i[(\Pi 2 x)_i - y\Pi 2 x], i = 1,2,...,n \quad (5)$$
A basic drawback of normal game theory is the assumption that all players interact globally. It is more realistic to study local games e.g. games on a lattice where players interact only with their nearest neighbors. Also there are several modifications for game formulations.

The prisoner's dilemma game (PD) is a symmetric game where each player has two possible strategies cooperate C or defect D with payoff matrix

$$\Pi = \begin{bmatrix} R & S \\ T & U \end{bmatrix}, \text{ T>R>U>S, 2R>T+S} \qquad (6)$$

The stable solution of this game is for both players to defect.

In the games under consideration, players belong to one of two sets the stronger (more successful in the competition at time t) players and the weaker (at time t) ones from which a player will be chosen to be eliminated. It is important to realize that both sets change with time hence a player in the first set at time t may be in the second set at a later time t1>t. Using PD to study this game one gets three cases: If the two players are safe from elimination then cooperation is highly likely since this is and iterated PD game [Hofbauer and Sigmund 1998]. If both players are unsafe then defect is the dominant strategy. If one player is safe and the other is not and since only the safe player can decide, then choosing the strongest player in the second set is the dominant strategy. This agrees with several observations in several TV programs which can be considered as experiments for this game. Hence we reach the following conclusion:

"In the abovementioned game if each player has a positive probability that he (she) will be in the set of weaker players (from which a player will be eliminated) at some time t>0 then it is likely that the overall winner of the game may not be the best player".

**3. Conclusions:**

In this paper we argue that in real life (or close to it) the statement "survival is for the fittest" is inappropriate. Our argument depends on the following points: First, almost every real problem is stochastic [Erdi 2008]. Hence the deductive statement "If A then B" should be replaced by "If A then B is highly likely to occur". Second biological fitness is a multi-objective concept hence the word "fittest" should be replaced by "non-dominated fitness". Consequently it is proposed to replace "survival is for the fittest" by "survival is for those with non-dominated fitness".
The results of abovementioned TV games support our conclusion.

References:
Ehrgott M. (2005), "Multi-criteria Optimization", Springer.
Erdi P. (2008), "Complexity explaied", Springer.
Hofbauer J. and Sigmund K. (1998), " Evolutionary games and population dynamics", Canbridge Univ. Press.
Webb J. (2007), "Game theory", Springer.